\documentclass[conference]{IEEEtran}
\usepackage{cite}
\usepackage{graphicx}
\usepackage{algorithmic}
\usepackage{enumitem}
\usepackage{url}
\usepackage{booktabs}
\usepackage{amsmath}
\usepackage{balance}
\usepackage{color}

\newcounter{example}[section]

\begin{document}
%
\title{Cross-status Communication and Project Outcomes in OSS Development--\\A Language Style Matching Perspective}

\author{\IEEEauthorblockN{Yisi Han$^\dagger$, Zhendong Wang$^\ddagger$, Yang Feng$^\dagger$, Zhihong Zhao$^\circ$, Yi Wang$^\ast$}
\IEEEauthorblockA{$^\dagger$Nanjing University, Nanjing, Jiangsu 210093, China\\$^\ddagger$University of California, Irvine, CA 92697, USA\\
$^\circ$Nanjing Tech University, Nanjing, Jiangsu 211816, China \\
$^\ast$Beijing University of Posts and Telecommunications, Beijing 100876, China\\
Email: $^\dagger$hanysannie@gmail.com, $^\ddagger$zhendow@uci.edu, $^\dagger$fengyang@nju.edu.cn, \\$^\circ$zhaozhih@njtech.edu.cn, $^\ast$yiwang@bupt.edu.cn}
}


%


\maketitle
\footnotetext[1]{This study was accepted at the MSR 2021 Registered Reports Track}
\begin{abstract}
Background. The success of an open source software (OSS) project requires effective communication among its members. Given that OSS projects often have established social status systems, much communication may happen between individuals of different statuses, particularly, elite developers who have project management privileges and ordinary project contributors. Since sociolinguistics literature and our prior work have found that groups in different status would be likely develop different language styles, which may hinder critical cross-status communication, and hereby influence the project’s outcomes.\\
Objectives. We seek to develop an understanding of cross-status communication, as well as its impacts on an OSS project’s outcomes in terms of productivity and quality. The anticipated results describe the linguistic similarities and differences of cross-status communication and reveal the relationships between these linguistic similarities and differences and project outcomes. \\
Method. We approach the above research objectives with the language style matching (LSM) tool which measures the similarities of cross-status communication in multiple language style features. For each sampled project, we first dynamically identify elite developers having project administration privileges. Then, we capture the cross-status communication between elite and non-elite developers; and calculate the LSM features of these two group of individuals. The LSM variables, together with project outcomes, will be used to fit regression models to analyze potential relationships between cross-status communication and project outcomes.\\
Limitations. The study has several limitations. First, it considers project hosted on GITHUB only. Second, to ensure data availability, our sample is drawn from top projects, thus may not be able to represent all projects. Third, we only consider a limited number of linguistic variables.  
\end{abstract}


%
\IEEEpeerreviewmaketitle

\section{Introduction}

It has long been known that the success of an open source project is largely determined by effective and efficient communication among its developers \cite{xuan2012measuring,kavaler2017perceived,vale2020relation,bettenburg2010studying,wolf2009predicting,el2019empirical}. Many of key technical and non-technical activities in open source development rely on interpersonal communication, to name a few, discussing issues, attracting new developers \cite{ducheneaut2005socialization,von2003community}, evaluating developers' contribution \cite{tsay2014influence, marlow2013impression} and active level \cite{xuan2016converging}. Moreover, there is convincing evidence to support Conway's law, which hypothesizes that ``a development organization will inevitably design systems that mirror its organizational communication structure'' \cite{conway1968committees}. Communication among developers in open source projects is contingent on many organizational and social factors \cite{fielding1999shared}. An important factor is the social status of developers \cite{scacchi2004free,stewart2005social}. In open source development, many key activities involve cross-status communication between developers of different statuses. For instance, entry-level, low-status developers often need to communicate their technical contributions with high-status, elite developers who hold project management privileges as defined in \cite{wang2020unveiling}. 


However, we have a very limited understanding of the cross-status communication in OSS development, and its impact on project outcomes. Sociolinguistic literature argues that people with different social statuses tend to develop different language use styles \cite{niederhoffer2002linguistic}, which may lead to potential communication barriers in cross-status communication \cite{steinmacher2015social,stewart2005social}. Meanwhile, social status enables the formation of organizational routines, guaranteeing a bottom-line function of communication among developers \cite{barker2010strategic}. The tension between these two forces may result in complicated dynamics in cross-status communication, which is critical to many important tasks in OSS development, to name a few, pull-request evaluation, bug-fixing, etc.

Thus, we argue that there is an imperative to investigate cross-status communication in OSS development. Doing so will enable us to design effective tools and mechanisms to support cross-status communications in OSS development. As we know, communication includes two parts, what and how people say \cite{kavaler2017perceived}. Language style matching (LSM) \cite{gonzales2010language}, defined as language use similarity on various linguistic word use features, provides a convenient yet powerful tool to study the cross-status communication in OSS development. With it, we would be able to evaluate the similarities and differences between elite and non-elite developers' language uses in their cross-status communications using data from multiple projects. Moreover, researchers have demonstrated LSM as a strong indicator of the outcomes of many social processes such as cooperation, attraction, cohesion, and group performances, in multiple domains (see Section II.B \& C).   

From a language style matching perspective, this study proposes to investigate cross-status communication in OSS development and its relationships with project outcomes. We plan to collect, clean, and compile a dataset from 200 randomly sampled large OSS projects hosting their primary repositories on \textsc{GitHub} (\textbf{Section VI}). Using this dataset, we are going to perform a series of analyses (\textbf{Section VII}) to (1) develop basic understandings of LSM in OSS projects (exploratory study), and (2) reveal the correlations between LSM and project outcomes (confirmatory study). 

\noindent \emph{Paper Organization.} The rest of this paper proceeds as follows. Section II briefly introduces the background of the study, including some basics about LSM and communication in OSS projects. Section III presents the research questions and corresponding hypotheses. Section IV through VII form the standard components of a registered report, including variables, materials, execution plan, and analysis plan. In section VIII, we append a brief statement on publishing generated data and other study materials.

\section{Background}

\subsection{Core Developers vs. Elite Developers}

Traditionally, members' roles in an open source project are often referred as ``core" and ``peripheral'' developers. Crowston et al. \cite{2006Core} define core developers as those `` who contribute most of the code and oversee the design and evolution of the project.'' They are at the center of the onion-like structure. Such terminology is a dominant terminology used in SE literature to characterize developers' roles and relationships resulting from role divergences. Core developers are often identified by counting their contributions, though some recent work starts utilizing the network-based perspectives to find core developers having significant impacts in the developer network drawn from source code's dependencies \cite{joblin2017classifying}.

However, we argue that the ``core-peripheral'' terminology does not fit the current study. The core-peripheral based member classes could not capture the social status and corresponding power relationships in open source projects, particularly under the circumstance that projects with commercial sponsorship become prevalent \cite{germonprez2017theory}. It is not uncommon for a company to assign its employees to sponsored projects and guarantee them full project privileges. These employees are powerful even with limited technical contributions. Besides, it is also challenging to conceptualize and operationalize the dynamic nature of the roles \cite{crowston2008free}, particularly for the downward role migrations (core $\rightarrow$ peripheral). Once members are identified as the cores for their early contributions, there is no straightforward to determine if they still could be viewed as core developers unless there are some clear signals for turnover.  

Thus, we follow the ``elite'' and ``non-elite'' notion introduced in our prior work \cite{wang2020unveiling}. As we mentioned in the Introduction, elite developers are those who obtain project management privileges regardless of their technical contributions. Doing so enables us to have a better conceptualization of power relationships in a project. Moreover, it also brings significant benefits at the operationalizational level. We could easily identify the elite dynamically by observing their executions of project management activities in project repositories over time. 

\subsection{Language Style Matching}

Language style matching (LSM) is a technique assessing the stylistic similarities in language use across groups and individuals \cite{gonzales2010language}. For conversations between two parties (either individual speakers or aggregated groups), LSM calculates the similarity in the use of function words such as pronouns and articles. Thus, it provides a convenient way for us to study the cross-status conversations in open source development.

Calculating LSM relies on a specific word categorization system. One of the most adopted such systems in LSM calculation is LIWC (Linguistic Inquiry and Word Count \cite{pennebaker2001linguistic}). LIWC2015 could provide the percentage of total words in a text that falls into 8 basic categories of function words: \emph{Personal pronouns}, \emph{Impersonal pronouns}, \emph{Articles}, \emph{Prepositions}, \emph{Auxiliary verbs}, \emph{Common adverbs}, \emph{Conjunctions}, and \emph{Negations}. These percentages could be used to compute LSM \cite{ireland2011language}. Let us suppose there are two parties A vs. B (for example, in our study, the elites vs. the non-elites), their percentages on a specific category \emph{i} is $Cat_{iA}$ and $Cat_{iB}$, LSM on category \emph{i} can be calculated using the following formula:
\begin{equation}
    LSM_{i} = 1-\frac{|Cat_{iA}-Cat_{iB}|}{Cat_{iA}+Cat_{iB}+0.0001}
    \label{lsm}
\end{equation}
In this formula, 0.0001 is added to the the denominator of the right part to avoid the potential ``Divided by 0'' problem.

Obviously, the function word categories only deal with the syntactic aspect of language use rather than the semantic aspect of language use. To solve this issue, we use four summary variables in LIWC2015 to capture the language styles related to semantics. These four summary variables are: \emph{Analytical Thinking}, \emph{Clout}, \emph{Authentic}, and \emph{Emotional Tone}.

In total, there are 12 variables. Following the conventional method \cite{ireland2014language}, we average the 12 category-level LSM scores to create a composite LSM score. In our study, higher numbers represent the greater overall stylistic similarity between a project's elites and non-elites.

\subsection{LSM and Communication in OSS Projects}

Psychologists and sociolinguists have revealed that LSM is an important indicator of interpersonal and group mimicry \cite{chartrand1999the} in human communication \cite{ireland2011language}, and thus influences psychological factors and behavioral outcomes in many social domains. Tab. \ref{lsm} provides a small set of such examples. These studies, while varying in their scientific focuses, convey a pretty consistent message that high LSM would indicate a certain level of cooperation (or the willingness of cooperation) and may hereby lead to favorable outcomes. 

\begin{table}[!h]
    \centering
    \caption{Some examples of the relationships between LSM and other constructs. }
    \begin{tabular}{ll}
    \toprule
    \textbf{Domain}  & \textbf{Reference} \\
    \midrule
    Criminology     &  Richardson et al. (2014) \cite{richardson2014language}\\
    International Politics & Bayram \& Ta (2018) \cite{bayram2019diplomatic} \\
    Healthcare & Rains (2016) \cite{rains2016language}\\
    Group Communication & Gonzales et al. (2010) \cite{gonzales2010language}\\
    Social Networking & Kovacs \& Kleinbaum (2019) \cite{kovacs2020language} \\
    Romantic Relationship & Ireland et al. (2011) \cite{ireland2011language}\\
    \bottomrule
    \end{tabular}
    \label{lsm}
\end{table}

Meanwhile, organizations, no matter formal or informal ones, have the tendency to implement certain kinds of status systems. Members of different statuses are likely to develop different language styles, as social externalizations of establishing status enabled norms \cite{niederhoffer2002linguistic,danescu2011mark}. There is no exception for OSS projects \cite{liao2019status,scacchi2004free}. In most projects, power is often shared by a small number of members holding project management privileges--the so-called ``elite developers'' defined in Wang et al. \cite{wang2020unveiling}. Elite and non-elite developers are likely to have developed different language styles. 

As we mentioned in the Introduction, cross-status communication plays an important role in OSS development because it relates to many critical tasks in the development \cite{stewart2005social}. Thus, language style similarities and differences in cross-status communication might have a profound impact on the development process and eventually influence a project's outcomes. However, the relationship between cross-status communication and project outcome could be complicated. According to literature in LSM, similar language styles (high LSM) may indicate good cooperation, few barriers, and shared common ground, thus having some positive effects. However, in the organizational context, different language styles, to some degree, reflect certain social norms. Too high language styles similarities in cross-status communication may blur the cognition of social norms among involved parties and lead to unfavorable outcomes \cite{markowitz2018academy,wang2020the}. Moreover, given that simply using a composite LSM score in an empirical study may be too coarse-grained, LSM values of different word categories may also show varying effects \cite{cannava2017language}, further complicating the picture yet providing more opportunities to design possibilities for supporting cross-status communication.  

Therefore, it is necessary to empirically examine the relationship between cross-status communication and project outcomes from a language style matching perspective. Unfortunately, as far as our best knowledge, cross-status communication has not yet received much attention in software engineering research. The proposed study is exactly designed to fill this gap to develop such an understanding from a language style matching perspective. The insights gaining through the study would inform the design of tools and mechanisms in supporting cross-status communication in open source development.


\section{Research Questions \& Hypotheses}
Specifically, we are going to perform a study driven by the following two research questions.
\begin{description}[leftmargin=*]
\item[$\mathbf{RQ_1}$] \emph{What are the Language Style Matching (LSM) values in OSS projects' cross-status communication between elite and non-elite developers? How do such values vary across different projects in different categories?}
\end{description}

Given the exploitative nature of the $\mathbf{RQ_1}$, no predetermined hypothesis is needed for this research question.

\begin{description}[leftmargin=*]
\item[$\mathbf{RQ_2}$] \emph{What are the relationships between a project Language Style Matching (LSM) values in cross-status communication and its outcomes?}
\end{description}

Since the project outcomes are measured in productivity and quality following Wang et al. \cite{wang2020unveiling}, we formalize the two hypotheses (and corresponding null hypotheses) that we aim to test, described as follows.

\begin{itemize}[leftmargin=*]
\item A project's LSM values have significant correlations with its productivity. [$\mathbf{H1}$]
\item A project's LSM values do not have significant correlations with its productivity. [$\mathbf{H_{0}1}$]
\item A project's LSM values have significant correlations with its quality. [$\mathbf{H2}$]
\item A project's LSM values do not have significant correlations with its quality. [$\mathbf{H_{0}2}$]
\end{itemize} 

Note that we do not explicitly state the (positive or negative) directions of the correlations. The argument in Section II.B demonstrates that the relationship could be complicated, particularly when we consider multiple linguistic word categories. The directions would be revealed along with the analysis. 

It is possible that there is no clear relationship between LSM and project outcomes identified. In the paper related to this registered report, such negative results must be reported faithfully in this case. Such results may result from that LSM only capture and conceptualize a partial reality of cross-status conversations between the elite and non-elite developers. In this case, we need to seek other mechanisms to characterize and quantify the cross-status conversations in future work.

\section{Variables}
There are three classes of variables involved in the proposed study.
The first classes of variables are those measures for \textbf{project outcomes}. We plan to reuse the four variables defined in Wang et al. \cite{wang2020unveiling}. They are listed in Tab. \ref{dv} with brief explanations of their meanings. Using two variables for each outcome would help avoid the biases caused by the single measurement. These variables are the \emph{dependent variables} of the study. They represent multiple aspects of project outcomes which are multifaceted in nature. $\overline{NewC}$ is a widely used productivity indicator for OSS project \cite{vasilescu2015quality}. $\overline{BCT}$ shows how productive a project in dealing with bugs \cite{10.1145/1137983.1138027}. $\overline{BCT}$ is a straightforward indicator of quality. $\overline{BFR}$ one of the key metrics related to the defect removal process by characterizing the birth (find a new bug) to death (fix an existing bug) process of the defect removal \cite{levendel1990reliability}.   

\begin{table}[!h]
    \centering
    \caption{A brief summary of the four project outcome variables.}
    \begin{tabular}{llp{0.58\columnwidth}}
    \toprule
    \textbf{Outcome} & \textbf{Variable} & \textbf{Meaning}\\
    \midrule
    Productivity     & $\overline{NewC}$ & The average of the numbers of monthly new commits in the 36 months.\\
         & $\overline{BCT}$ & The average cycle time of a project’s closed bugs in a 36 months.\\
    \midrule
    Quality     & $\overline{NewB}$ & The average of the numbers of monthly newly found bugs in the 36 months.\\
         & $\overline{BFR}$ & The average of the ratios of monthly fixed bugs to newly found bugs in the 36 months.\\
    \bottomrule
    \end{tabular}
    
    \label{dv}
    \vspace{-1em}
\end{table}

The second class contains the \textbf{LSM variables}, measured on corresponding LIWC categories. We plan to use 12 such variables and one composite variable (8 function words + 4 summary, see Section II.B). All of them are calculated with the formula \ref{lsm}. The \emph{independent variables} of the study would be drawn from them.

In addition to the above variables, the study would incorporate six \textbf{control variables}. The first control variable is the \emph{average ratio of the elites to the non-elites}. The second is the \emph{project size}. The third is the \emph{company sponsorship}, a binary variable capturing if commercial companies sponsor the project. The fourth one is the \emph{average experience of a project's developers} on \textsc{GitHub} at the time of the project starts. The fifth is the \emph{main programming language} used by a project. The last one is the \emph{project's domain}. These two are categorical variables. The categories would be determined after sampling the targeted projects. 

\section{Materials}
\subsection{Dataset} 
We intend to use the data collected from a sample of 200 top software development projects hosted on \textsc{GitHub}. The sampling of these projects is introduced in the execution plan. For each project, the data contains textual conversations that happen between elite developers and non-elite developers logged in the first 36 months of the development process. The data also contains all the events in the project's repository, which enables us to extract project outcomes. Project metadata is also included in the dataset.  

\subsection{Software}
There are two types of software tools that will be used in this study. The first set of tools are used to collect and prepare data. Because the data are obtained from two data sources (\textsc{GitHub} and \textsc{GH Archive}, see section VI.A.1), we have to use both official \textsc{GitHub} API and customized queries written in \texttt{Python} to collect the data. In data preparation, the study has to deal with developers' conversations, so we use \texttt{Python}'s \texttt{NLTK} package to perform standard preprocessing. Besides, given these conversations' software engineering nature, we would also have to write \texttt{Shell} scripts with regular expressions in data preprocessing. Finally, computing LSM variables would use the LIWC2015 software\footnote{http://liwc.wpengine.com/} and dictionaries purchased from the Pennebaker Conglomerates, Inc.

The second type of software tools is those for statistical analysis. We choose the \texttt{R} statistical software (ver. 4.0.3) to perform descriptive analyses for answering $\mathbf{RQ_1}$ and build statistical models for answering $\mathbf{RQ_2}$. In addition to the standard \texttt{R} base environment, we also use a few other \texttt{R} packages such as \texttt{glm} (for building regression models), \texttt{ggplot2} (for data visualizations) for the analyses specified in section VI.

\subsection{Hardware}
We expect to run the study with a high-performance work station in data collection and preparation and use several PC and Mac laptops to analyze the data and draft the report. When necessary, the dataset (or part of it) may be stored in an encrypted portable hard drive.

\section{Execution Plan}
We will follow the following execution plan to conduct the study. It consists of three major steps as follows.

\subsection{Step 1. Data Collection and Clean}
\subsubsection{Sampling Targeted Projects}
The sampling process starts with a list of 2,000 most starred software development projects primarily hosted on \textsc{GitHub} (excluding mirrors). These projects should also satisfy four following criteria: (1) adopting the pull-request contribution solicitation model, (2) establishing elite vs. non-elite status, (3) having at least 36 months of project history, and (4) having traceable records of continuous
contributions from a set of contributors (at least 100 pull-requests and 50 contributors historically). From this list, we randomly draw a sample of 200 projects. 

\subsubsection{Data Collection}
For each project, we plan to collect data from two sources: \textsc{GitHub} API and \textsc{GH Archive}\footnote{Access via Google BigQuery: \url{https://console.cloud.google.com/bigquery?project=githubarchive&page=project&pli=1}.}. We use two sources because the official \textsc{GitHub} API has put limitation access to a project's historical event \footnote{The current version of the \textsc{GitHub} API only allows to access 300 events or events in up to
the past 90 days, whichever is met first.}.  
Tab. \ref{data} summarizes the different data we collected from the above two sources. As shown in the table, conversations are collected using \textsc{GitHub} API. The conversations would include comments around all issues, pull requests, and commits.
\begin{table}[!h]
    \centering
    \caption{Different types of data collected from two data sources}
    \begin{tabular}{ll}
    \toprule
      \textbf{Data Source}   &  \textbf{Data Type} \\
    \midrule
    \textsc{GitHub} API     & List of contributors and their basic information\\
                            & Conversations in multiple unit of the project repository  \\
                            & Repository metadata\\
    \midrule
    \textsc{GH Archive}      & Public project events \\
    \bottomrule     
    \end{tabular}
    
    \label{data}
\end{table} 

\subsubsection{Data Preprocessing}
The data has to be preprocessed for later operations. First, we keep each project's first 36 months' data only. Doing so would help mitigate the potential problem caused by collecting data from different phases in a project's life cycle. Second, we remove the redundant event data caused by retrieving data from two sources and resolving the entities to link the data from the same individual. All event data is organized on a monthly basis. 

We now come to the main task of data processing--cleaning the textual conversations. Given that all these conversations are related to software engineering, we need to first deal with some SE-specific textual content, for example, removing the hashcode embedded in these conversations. We plan to perform two sub-tasks. First, we compile a dictionary containing common acronyms in software engineering and use it to replace these acronyms with full words. Second, we plan to develop a \texttt{shell} script consisting of a set of regular expressions. After finishing these procedures, standard text preprocessing libraries such as \texttt{Python}'s \texttt{NLTK} would take the remainder of text data preprocessing work. Note that some special information that would be helpful in determining cross-status communication is retained in this step, for instance, ``@-mentions'' in multiple-individual conversations.

\subsection{Step 2. Data Preparation}
\subsubsection{Identifying Elite Developers}
As mentioned in the Introduction, we use ``elite'' vs. ``non-elite'' to reflect the status differences among developers, where the elites are those who have project management privileges. The identification of these elite developers is performed by using the method proposed in Wang et al. \cite{wang2020unveiling} based on a project's fine-grained events. To be specific, if a developer performs any action requiring the ``write'' permission, we mark him/her as an elite developer. However, being an elite is not permanent. If that developer does not perform any action requiring the write permission, his/her elite status would not be renewed; otherwise, the status would be extended for another three months. This method well reflects the dynamic nature of an individual's status in a project.

Note that setting the three months window opens some possibility of potential bias since developers who lose their elite status might still be counted as elite for at most three months. The three-month window designation in determining elite status is based on our empirical observations and prior literature. Most cases that developers lose their elite status result from these developers' turnovers. There are varying practices in SE literature about the threshold for determining turnovers. Foucault et al. uses 180 days \cite{10.1145/2786805.2786870}; Lin et al. experiments with 180 days, 90 days, 30 days \cite{lin2017developer}. We choose three months (~90 days) because there are cases that 30 days duration was not enough (particularly in vacations and holiday seasons), and 90 days provide almost identical results as 180 days. Besides, since the turnover causes most cases of losing elite status, those developers rarely participate in conversations after that. Therefore, the potential bias should be very minimal.

\subsubsection{Identifying Conversations Involving Both Elite and Non-Elite Developers}
This study's primary goal requires us to focus on cross-status communication rather than all conversations in a project. We need to filter out the conversations happening between people of the same status. Using the list of identified elite developers, we plan to examine every conversation to check if it involves both elite and non-elite developers; otherwise, we exclude it. In case of a conversation involving multiple developers, we would leverage the fine-grained \emph{replying} history of the conversation thread to determine cross-status conversations. We first check the speaker's status who initiates the conversation; we would remove the replies from those who share the same status, unless they have specific @-mentions to the individuals of different status.

Because the analysis unit of this study is at the aggregated project level, the scope of the study has little to do with specific conversations. Therefore, once all cross-status conversations are identified, we create two plain text files for each project. Each file contains all aggregated text of speakers having the same status. We use these two text files to extract the scores of LSM variables for each project. 

In some rare cases, a conversation's participant may experience role changes during the conversation. Since the study focuses on group-level differences, we would not deal with individual changes of status explicitly. For example, let us suppose in a conversation, at time 1, A who was not an elite said something; later, he or she said something at time 2 when the status had changed to an elite. In our study, we put A's words in time 1 and 2 separately. I.e., A's words at time 1 would go to the collection of the non-elites' words; and A's words at time 2 would go to the collection of the elites' words. Indeed such cases are very rare in empirical data. Given the overwhelmingly large amount of data and the focus on the statistics at the group-level, we thus argue such status changes within conversations are not likely to cause potential issues.

\subsubsection{Extracting Variables' Values}

Computing the values for the four project outcome variables takes two steps. For each project, we first calculate the values of these four variables in each month, and then compute the average of them as the final scores of project outcomes of each project in the entire 36 months. For project $i$, we use $<\overline{NewC_i}, \overline{BCT_i}, \overline{NewB_i}, \overline{BFR_i}>$ to represent them. Calculating the last three variables requires distinguishing bug-related issues from other issues. We search a list of bug-related keywords (e.g., defect, error, bug) in every issue's tags and titles. If any one of these keywords appears, the corresponding issue would be identified as a bug issue. 

Once we obtain a project's all conversational data between elite and non-elite, calculating the scores of LSM variables is straightforward. Because the analysis unit of the study is at the project level, we would focus on aggregated LSM rather than individual level LSM. For each project, we separate the elites' and the non-elites' discourses in cross-status conversations, and compute the LIWC scores for both of them. Then, we use formula (1) to calculate the LSM values in the 12 categories. Conventionally, the composite LSM would be generated by averaging the 12 categorical LSMs \cite{ireland2011language}. For project $i$, we use $<LSM_{0i}, LSM_{1i}, ..., LSM_{12i}>$ to represent them, where $LSM_0i$ represents the composite LSM. 

The identification of elite developers allows us to conveniently compute our first control variable, the average ratio of the elites to the non-elites. Project size would be represented by the average number of all developers. We plan to manually check if a project is sponsored by any commercial companies. We would calculate the average experience of a project's contributors. A project's main programming language and domain could be fairly easy to obtain through manual coding procedures. For project $i$, we use $<Control_1, Control_2, Control_3, Control_4, Control_5, \\Control_6>$ to represent them. Categorical variables $Control_5$ and $Control_6$ have to be recorded when using them in regressions (see Section VII.B). 

We then have all values for the project $i$, represented by a 23-tuple by joining $<\overline{NewC_i}, \overline{BCT_i}, \overline{NewB_i}, \overline{BFR_i}>$; $<LSM_{0i}, LSM_{1i}, ..., LSM_{12i}>$; as well as $<Control_1,\\ Control_2, Control_3, Control_4, Control_5, Control_6>$. Since we have 200 sampled projects, there are 200 such 23-tuples. We store them into a \texttt{.csv} file for data analysis.

\subsection{Step 3. Data Analysis}
In this step, we perform data analysis to gain insights about the language style matching of elite and non-elite developers ($\mathbf{RQ_1}$), as well as the relationships between them and project outcomes $\mathbf{RQ_2}$. In the following analysis plan (Section VI), we provided detailed descriptions of how we use proper statistical techniques to answer the two research questions.

\section{Analysis Plan}
\subsection{Analysis Plan for $\mathbf{RQ_1}$}
Answering $\mathbf{RQ_1}$ requires us to perform a series of descriptive statistical analyses. We plan to first calculate the values of all LSM variables for each project. We would then combine all projects' LSM values to study their distributions and perform cross-project and cross-category comparisons. We plan to visualize them to facilitate the understanding and reasoning of LSM values and their differences intuitively. These visualizations would include but are not limited to box plots, histograms, kernel density estimations, scatter charts, and so on.    

In addition to the above descriptive analyses on LSMs of cross-status conversations, we would also analyze within-status conversations. Doing so enables us to determine if the variances in language styles result from status differences or merely the common variances in any interpersonal communication among team members. Since there is no way to obtain LSM values for within-status conversations, we plan to calculate and compare the selected LIWC categories' raw values in three corpora: cross-status conversations, within-status conversations among the elites, and within-status conversations among the non-elites. It would partially inform us about the sources of the variances in language styles. The results will be reported accordingly.

\subsection{Analysis Plan for $\mathbf{RQ_2}$}
Answering $\mathbf{RQ_2}$ requires examining the relationships between LSM variables and project outcomes. It is straightforward to employ Ordinary Least Squared (OLS) regression \cite{wooldridge2016introductory} with all data cases in our sample to fulfill this task. We build four regression models for each project outcome variable. In these four models, the \emph{dependent variable} is one of the project outcome variables. However, the \emph{independent variables} are four different sets of LSM variables: (1) the composite LSM variable, (2) the 8 LSM variables of function words, (3) the 4 LSM summary variables, and (4) all 12 category-level LSM variables. Thus, we have 16 models in total. We also include the control variables, including the average ratio of the elites to the non-elites, size, company sponsorship, average experience, the project's main programming language, and its domain. Take as the $\overline{NewC}$ an example, the corresponding regression model with 8 LSM variables of function words is in the following form:
\begin{equation}
    \overline{NewC} = \beta_{0} + \sum_{j=1}^{8}\beta_{j} \times LSM_{j} + \sum_{k=9}^{m}\beta_{k} \times Control_{k-8} + u\\
\end{equation}
Note that categorical variables having more than two categories cannot directly enter the regression equations. They have to be represented by a series of dummy variables. We do not know exactly the number of dummy variables needed until sampling the project. Therefore, we use a ``$m$'' to represent the total number of independent and control variables in the above equation.

For each project outcome variable, a model containing control variables only is going to be built as the benchmark for model comparisons. Therefore, we have 20 models in total. Standard model diagnostics for OLS regressions, for example, linearity, homogeneity, normality, and leverage cases, would be performed in the model building process. Besides, the possible multiple collinearity resulting from correlated independent variables would also be treated in the model building process using VIF tests.

In addition to the above general model diagnostics, there are potential nonlinear effects between LSM and project outcomes. For example, when LSM is low, the project outcomes may not be good; but if LSM is too high, the project outcomes may not be good either due to the absence of social norms (inverse U-shaped relationships). We would do supplemental analysis to the models built with the composite LSM variable by adding its squared into the regressions to examine the potential quadratic effects. 

\section{Publication of Generated Data and Other Study Materials}

All generated data in the research process will be shared in a DOI citable long-term archive with the MIT license. Other study materials, including but not limited to, detailed logs of study execution, scripts for data collection, preparation, and analysis, will also be published along with the generated data as the study's replication package.  

\section{Potential Implications}

Establishing the relationships between the LSM metrics and project outcomes could bring immediate implications to future research and practices. For research, it demonstrates the importance of investigating cross-status communication among open source developers and opens rich future research opportunities. For example, what are the detailed mechanisms of cross-status communication influence project outcomes? Are there any other social factors involved? How to drill down to individual conversations? For OSS development practices, such relationships could inform decision-makers in OSS projects to reflect on how to facilitate cross-status conversations to achieve better project outcomes, as well as identify potential problems existing in current cross-status conversations. In case no relationship between the LSM metrics and project outcomes could be established, the study also has potential implications. It would at least demonstrate that LSM perhaps not performs well in conceptualizing and quantifying cross-status conversations and urge researchers, including ourselves, to explore other alternatives.

\balance
\bibliographystyle{IEEEtran}
\bibliography{sample}

\begin{thebibliography}{10}
\providecommand{\url}[1]{#1}
\csname url@samestyle\endcsname
\providecommand{\newblock}{\relax}
\providecommand{\bibinfo}[2]{#2}
\providecommand{\BIBentrySTDinterwordspacing}{\spaceskip=0pt\relax}
\providecommand{\BIBentryALTinterwordstretchfactor}{4}
\providecommand{\BIBentryALTinterwordspacing}{\spaceskip=\fontdimen2\font plus
\BIBentryALTinterwordstretchfactor\fontdimen3\font minus
  \fontdimen4\font\relax}
\providecommand{\BIBforeignlanguage}[2]{{%
\expandafter\ifx\csname l@#1\endcsname\relax
\typeout{** WARNING: IEEEtran.bst: No hyphenation pattern has been}%
\typeout{** loaded for the language `#1'. Using the pattern for}%
\typeout{** the default language instead.}%
\else
\language=\csname l@#1\endcsname
\fi
#2}}
\providecommand{\BIBdecl}{\relax}
\BIBdecl

\bibitem{xuan2012measuring}
Q.~Xuan, M.~Gharehyazie, P.~T. Devanbu, and V.~Filkov, ``Measuring the effect
  of social communications on individual working rhythms: A case study of open
  source software,'' in \emph{Proc. Socialinfo'12}.\hskip 1em plus 0.5em minus
  0.4em\relax IEEE, 2012, pp. 78--85.

\bibitem{kavaler2017perceived}
D.~Kavaler, S.~Sirovica, V.~Hellendoorn, R.~Aranovich, and V.~Filkov,
  ``Perceived language complexity in github issue discussions and their effect
  on issue resolution,'' in \emph{Proc. ASE'17)}, 2017, pp. 72--83.

\bibitem{vale2020relation}
G.~Vale, A.~Schmid, A.~R. Santos, E.~S. De~Almeida, and S.~Apel, ``On the
  relation between github communication activity and merge conflicts,''
  \emph{Empirical Software Engineering}, vol.~25, no.~1, pp. 402--433, 2020.

\bibitem{bettenburg2010studying}
N.~Bettenburg and A.~E. Hassan, ``Studying the impact of social structures on
  software quality,'' in \emph{Proc. ICPC'10}, 2010, pp. 124--133.

\bibitem{wolf2009predicting}
T.~Wolf, A.~Schroter, D.~Damian, and T.~Nguyen, ``Predicting build failures
  using social network analysis on developer communication,'' in \emph{Proc.
  ICSE'09}.\hskip 1em plus 0.5em minus 0.4em\relax IEEE, 2009, pp. 1--11.

\bibitem{el2019empirical}
M.~El~Mezouar, F.~Zhang, and Y.~Zou, ``An empirical study on the teams
  structures in social coding using github projects,'' \emph{Empirical Software
  Engineering}, vol.~24, no.~6, pp. 3790--3823, 2019.

\bibitem{ducheneaut2005socialization}
N.~Ducheneaut, ``Socialization in an open source software community: A
  socio-technical analysis,'' \emph{Computer Supported Cooperative Work
  (CSCW)}, vol.~14, no.~4, pp. 323--368, 2005.

\bibitem{von2003community}
G.~Von~Krogh, S.~Spaeth, and K.~R. Lakhani, ``Community, joining, and
  specialization in open source software innovation: a case study,''
  \emph{Research policy}, vol.~32, no.~7, pp. 1217--1241, 2003.

\bibitem{tsay2014influence}
J.~Tsay, L.~Dabbish, and J.~Herbsleb, ``Influence of social and technical
  factors for evaluating contribution in github,'' in \emph{Proc. ICSE'14},
  2014, pp. 356--366.

\bibitem{marlow2013impression}
J.~Marlow, L.~Dabbish, and J.~Herbsleb, ``Impression formation in online peer
  production: activity traces and personal profiles in github,'' in \emph{Proc.
  CSCW'13}, 2013, pp. 117--128.

\bibitem{xuan2016converging}
Q.~Xuan, P.~Devanbu, and V.~Filkov, ``Converging work-talk patterns in online
  task-oriented communities,'' \emph{PLOS One}, vol.~11, no.~5, p. e0154324,
  2016.

\bibitem{conway1968committees}
M.~E. Conway, ``How do committees invent,'' \emph{Datamation}, vol.~14, no.~4,
  pp. 28--31, 1968.

\bibitem{fielding1999shared}
R.~T. Fielding, ``Shared leadership in the apache project,''
  \emph{Communications of the ACM}, vol.~42, no.~4, pp. 42--43, 1999.

\bibitem{scacchi2004free}
W.~Scacchi, ``Free and open source development practices in the game
  community,'' \emph{IEEE software}, vol.~21, no.~1, pp. 59--66, 2004.

\bibitem{stewart2005social}
D.~{Stewart}, ``Social status in an open-source community,'' \emph{American
  Sociological Review}, vol.~70, no.~5, pp. 823--842, 2005.

\bibitem{wang2020unveiling}
Z.~Wang, Y.~Feng, Y.~Wang, J.~A. Jones, and D.~Redmiles, ``Unveiling elite
  developers’ activities in open source projects,'' \emph{ACM Transactions on
  Software Engineering and Methodology (TOSEM)}, vol.~29, no.~3, pp. 1--35,
  2020.

\bibitem{niederhoffer2002linguistic}
K.~G. Niederhoffer and J.~W. Pennebaker, ``Linguistic style matching in social
  interaction,'' \emph{Journal of Language and Social Psychology}, vol.~21,
  no.~4, pp. 337--360, 2002.

\bibitem{steinmacher2015social}
I.~Steinmacher, T.~Conte, M.~A. Gerosa, and D.~Redmiles, ``Social barriers
  faced by newcomers placing their first contribution in open source software
  projects,'' in \emph{Proc. CSCW'15}, 2015, pp. 1379--1392.

\bibitem{barker2010strategic}
R.~T. Barker and K.~Gower, ``Strategic application of storytelling in
  organizations: Toward effective communication in a diverse world,'' \emph{The
  Journal of Business Communication (1973)}, vol.~47, no.~3, pp. 295--312,
  2010.

\bibitem{gonzales2010language}
A.~L. Gonzales, J.~T. Hancock, and J.~W. Pennebaker, ``Language style matching
  as a predictor of social dynamics in small groups,'' \emph{Communication
  Research}, vol.~37, no.~1, pp. 3--19, 2010.

\bibitem{2006Core}
K.~Crowston, K.~Wei, Q.~Li, and J.~Howison, ``Core and periphery in free/libre
  and open source software team communications,'' in \emph{Proc. HICSS '06},
  2006, pp. 118:1--10.

\bibitem{joblin2017classifying}
M.~Joblin, S.~Apel, C.~Hunsen, and W.~Mauerer, ``Classifying developers into
  core and peripheral: An empirical study on count and network metrics,'' in
  \emph{Proc. ICSE'17}, 2017, pp. 164--174.

\bibitem{germonprez2017theory}
M.~Germonprez, J.~E. Kendall, K.~E. Kendall, L.~Mathiassen, B.~Young, and
  B.~Warner, ``A theory of responsive design: A field study of corporate
  engagement with open source communities,'' \emph{Information Systems
  Research}, vol.~28, no.~1, pp. 64--83, 2017.

\bibitem{crowston2008free}
K.~Crowston, K.~Wei, J.~Howison, and A.~Wiggins, ``Free/libre open-source
  software development: What we know and what we do not know,'' \emph{ACM
  Computing Surveys (CSUR)}, vol.~44, no.~2, pp. 1--35, 2008.

\bibitem{pennebaker2001linguistic}
J.~W. Pennebaker, M.~E. Francis, and R.~J. Booth, ``Linguistic inquiry and word
  count: Liwc 2001,'' \emph{Mahway: Lawrence Erlbaum Associates}, vol.~71, no.
  2001, p. 2001, 2001.

\bibitem{ireland2011language}
M.~E. Ireland, R.~B. Slatcher, P.~W. Eastwick, L.~E. Scissors, E.~J. Finkel,
  and J.~W. Pennebaker, ``Language style matching predicts relationship
  initiation and stability,'' \emph{Psychological Science}, vol.~22, no.~1, pp.
  39--44, 2011.

\bibitem{ireland2014language}
M.~E. Ireland and M.~D. Henderson, ``Language style matching, engagement, and
  impasse in negotiations,'' \emph{Negotiation and conflict management
  research}, vol.~7, no.~1, pp. 1--16, 2014.

\bibitem{chartrand1999the}
T.~L. {Chartrand} and J.~A. {Bargh}, ``The chameleon effect: The perceptional
  behavior link and social interaction.'' \emph{Journal of Personality and
  Social Psychology}, vol.~76, no.~6, pp. 893--910, 1999.

\bibitem{richardson2014language}
B.~H. Richardson, P.~J. Taylor, B.~Snook, S.~M. Conchie, and C.~Bennell,
  ``Language style matching and police interrogation outcomes.'' \emph{Law and
  Human Behavior}, vol.~38, no.~4, p. 357, 2014.

\bibitem{bayram2019diplomatic}
A.~B. Bayram and V.~P. Ta, ``Diplomatic chameleons: Language style matching and
  agreement in international diplomatic negotiations,'' \emph{Negotiation and
  Conflict Management Research}, vol.~12, no.~1, pp. 23--40, 2019.

\bibitem{rains2016language}
S.~A. Rains, ``Language style matching as a predictor of perceived social
  support in computer-mediated interaction among individuals coping with
  illness,'' \emph{Communication Research}, vol.~43, no.~5, pp. 694--712, 2016.

\bibitem{kovacs2020language}
B.~Kovacs and A.~M. Kleinbaum, ``Language-style similarity and social
  networks,'' \emph{Psychological Science}, vol.~31, no.~2, pp. 202--213, 2020.

\bibitem{danescu2011mark}
C.~Danescu-Niculescu-Mizil, M.~Gamon, and S.~Dumais, ``Mark my words!
  linguistic style accommodation in social media,'' in \emph{Proc. WWW'11},
  2011, pp. 745--754.

\bibitem{liao2019status}
J.~Liao, G.~Yang, D.~Kavaler, V.~Filkov, and P.~Devanbu, ``Status, identity,
  and language: A study of issue discussions in github,'' \emph{PLOS One},
  vol.~14, no.~6, p. e0215059, 2019.

\bibitem{markowitz2018academy}
D.~M. {Markowitz}, ``Academy awards speeches reflect social status, cinematic
  roles, and winning expectations:,'' \emph{Journal of Language and Social
  Psychology}, vol.~37, no.~3, pp. 376--387, 2018.

\bibitem{wang2020the}
Y.~{Wang}, ``The price of being polite: politeness, social status, and their
  joint impacts on community q\&a efficiency,'' \emph{Journal of Computational
  Social Science}, pp. 1--22, 2020.

\bibitem{cannava2017language}
K.~{Cannava} and G.~D. {Bodie}, ``Language use and style matching in supportive
  conversations between strangers and friends,'' \emph{Journal of Social and
  Personal Relationships}, vol.~34, no.~4, 2017.

\bibitem{vasilescu2015quality}
B.~Vasilescu, Y.~Yu, H.~Wang, P.~Devanbu, and V.~Filkov, ``Quality and
  productivity outcomes relating to continuous integration in github,'' in
  \emph{Proc. ESEC/FSE'15}, 2015, pp. 805--816.

\bibitem{10.1145/1137983.1138027}
S.~Kim and E.~J. Whitehead, ``How long did it take to fix bugs?'' in
  \emph{Proc. MSR'06}, 2006, p. 173–174.

\bibitem{levendel1990reliability}
Y.~Levendel, ``Reliability analysis of large software systems: Defect data
  modeling,'' \emph{IEEE Transactions on Software Engineering}, vol.~16, no.~2,
  pp. 141--152, 1990.

\bibitem{10.1145/2786805.2786870}
M.~Foucault, M.~Palyart, X.~Blanc, G.~C. Murphy, and J.-R. Falleri, ``Impact of
  developer turnover on quality in open-source software,'' in \emph{Proc.
  ESEC/FSE'15}, 2015, p. 829–841.

\bibitem{lin2017developer}
B.~Lin, G.~Robles, and A.~Serebrenik, ``Developer turnover in global,
  industrial open source projects: Insights from applying survival analysis,''
  in \emph{Proc. ICGSE'17}, 2017, pp. 66--75.

\bibitem{wooldridge2016introductory}
J.~M. Wooldridge, \emph{Introductory Econometrics: A Modern Approach}.\hskip
  1em plus 0.5em minus 0.4em\relax Nelson Education, 2016.

\end{thebibliography}



\end{document}